\documentclass[english,aps,prd, twocolumn]{revtex4-1}
\usepackage[T1]{fontenc}
\usepackage[latin9]{inputenc}
\usepackage{color}
\usepackage{amsmath}
\usepackage{graphicx}
\usepackage{esint}
\usepackage{amsfonts,amssymb}

\makeatletter
\usepackage{slashed}
\usepackage[colorlinks,
            linkcolor=blue,
            anchorcolor=red,
            citecolor=green
            ]{hyperref}
\usepackage{cleveref} 

\definecolor{XQ}{rgb}{1,0.5,0}
\definecolor{XQ}{rgb}{0,0,0} 

\def\XQ#1{{\color{XQ}#1}}

\definecolor{ZM}{rgb}{0, 0,1}
\definecolor{ZM}{rgb}{0, 0, 0}
\def\ZM#1{{\color{ZM}#1}}

\definecolor{Reply}{rgb}{0.5,0.0,0.5}
\def\Reply#1{{\color{Reply}#1}}
\definecolor{Reply}{rgb}{0,0,0}

\def\PRLequal{\,{=}\,}
\newcommand\trick[1]{}

\makeatother
\usepackage{cancel}
\usepackage{babel}
\begin{document}
\title{Torsion, energy magnetization and thermal Hall effect}
\date{\today}
\author{Ze-Min Huang$^1$}
\author{Bo Han$^2$}
\author{Xiao-Qi Sun$^1$}
\email{xiaoqi20@illinois.edu}
\affiliation{$^1$Department of Physics and Institute for Condensed Matter Theory, University of Illinois at Urbana-Champaign,
1110 West Green Street, Urbana, Illinois 61801, USA}
\affiliation{$^2$Theory of Condensed Matter Group, Cavendish Laboratory, University of Cambridge, \\
J.~J.~Thomson Avenue, Cambridge CB3 0HE, United Kingdom}

\begin{abstract}
We study the effective action of  hydrostatic response to torsion in the absence of spin connections in gapped $\left(2+1\right)$-dimensional topological phases. 
\XQ{In previous studies, a torsional Chern-Simons term with a temperature-squared ($T^2$) coefficient was proposed as an alternative action to describe thermal Hall effect with the idea of balancing the diffusion of heat by a torsional field. However, the question remains whether this action leads to local bulk thermal response which is not suppressed by the gap. In our hydrostatic effective action, we show that the $T^2$ bulk term is invariant under variations up to boundary terms considering the back reaction of the geometry on local temperature, which precisely describes the edge thermal current.}
\XQ{Furthermore, there is no boundary diffeomorphism anomalies and bulk inflow thermal currents at equilibrium and therefore no edge-to-edge adiabatic thermal current pumping. These results are in consistent with exponentially suppressed thermal current for gapped phases. } 
\end{abstract}
\maketitle

\section{Introduction}
The Chern-Simons term originating from external electromagnetic fields
is known to be the effective action for the integer quantum Hall effect, where the quantization of Hall conductance is guaranteed by gauge invariance~\citep{Laughlin:1981}, or boundary gauge anomalies~\citep{callan1985npb,Wen:1991,stone1991aop}.
Based on Luttinger's seminal work \citep{luttinger1964pr},
a torsional field has been introduced to balance the diffusion of
heat \citep{shitade2014ptep,gromov2015prl,bradlyn2015prb}. Analogously,
torsional Chern-Simons terms have been proposed to be the effective action
for torsional viscosity at zero temperature~\citep{hughes2011prl,hughes2013prd,onkar2014prd}
and thermal Hall effect at finite temperature ~\citep{shitade2014ptep,nakai2016njp,nakai2017prb,huang2020prb1,liang2020prr}.
However, torsional anomalies are controversial because of its dependence
upon ultra-violet (UV) cut-off~\citep{chandia1997prd,chandia2001prd,kreimer2001prd, obukhov1997fop, peeters1999jhep, yajima1996cqg, soo1999prd, nissinen2020prl, huang2020prb}.
A clear physical meaning for torsional anomalies
is thus highly needed. 

Recently, thermal Hall effect has been observed experimentally in gapped topological phases~\citep{Banerjee:2017,banerjee2018nature,Dutta:2021} 
and has now attracted much attention due to the observed large signature from charge neutral excitations~\citep{onose2010science,hirschberger2015science, Ideue:2017,kasahara2018nature, kasahara2018prl,grissonnanche2019nature,li2020prl,grissonnanche2020natphys,Yokoi:2021}. 
\XQ{However, despite of the fast evolving experimental techniques, the fundamental understanding of whether thermal Hall current flows through the bulk of these systems is still incomplete.} 
For gapped topological phases, based on anomaly matching and generalized Laughlin's
argument, it was suggested in Refs.~\citep{nakai2016njp,nakai2017prb}
that there can exist a bulk thermal Hall current. 
This argument contradicts results in Refs.~\citep{stone2012prb,bradlyn2015prb,yuval2019prb},
where bulk thermal Hall currents are always exponentially suppressed
by the bulk gap. Hence, we aim to resolve this contradiction here, which shall \XQ{add a new perspective} to investigate thermal Hall effect. 

In this paper, by coupling matter fields to teleparallel gravity,
we study the response of the matter to inhomogeneous gravitational field at equilibrium. A hydrostatic effective action is derived,
which turns out to be the torsional Chern-Simons term and its coefficient
is the energy magnetization~\citep{Cooper1997,qin2011prl,zhang2020prb}. Similar to torsional
Chern-Simons terms, energy magnetizations can contain a constant UV dependent piece at zero temperature for a continuous model such as the massive Dirac fermion~\footnote{See Ref.~\cite{guo2020prb} and also in our Appendix~\ref{subsec:THE_2+1}}. 
We further show that there can be a temperature-squared term in gapped systems. In sharp contrast with the zero temperature piece, this term can be recasted as a topological $\theta$-term in terms
of Kaluza-Klein gauge fields, such that it is invariant under variations
of background fields and manifests itself as boundary currents. Therefore, adiabatic change of the background field cannot induce a bulk thermal currents. 
 Also, from the boundary picture, the resulting boundary energy current does not possess diffeomorphism anomalies,
hence no bulk inflow energy current is needed to absorb boundary anomalies. However, boundary global gravitational anomalies
do quantize the change of the coefficient of this $\theta$-term across the boundary, which reveals the relative topological meaning~\cite{kapustin2020prb} of the $\theta$-term between adjacent materials. Apart from addressing the described debates, our theory provides a top-down approach for the magnetization and energy magnetization: we show that various properties of the magnetization and energy magnetization can be obtained from macroscopic effective action with symmetry considerations and are independent of details of the microscopic model.

 \begin{figure}
\includegraphics[scale=0.33]{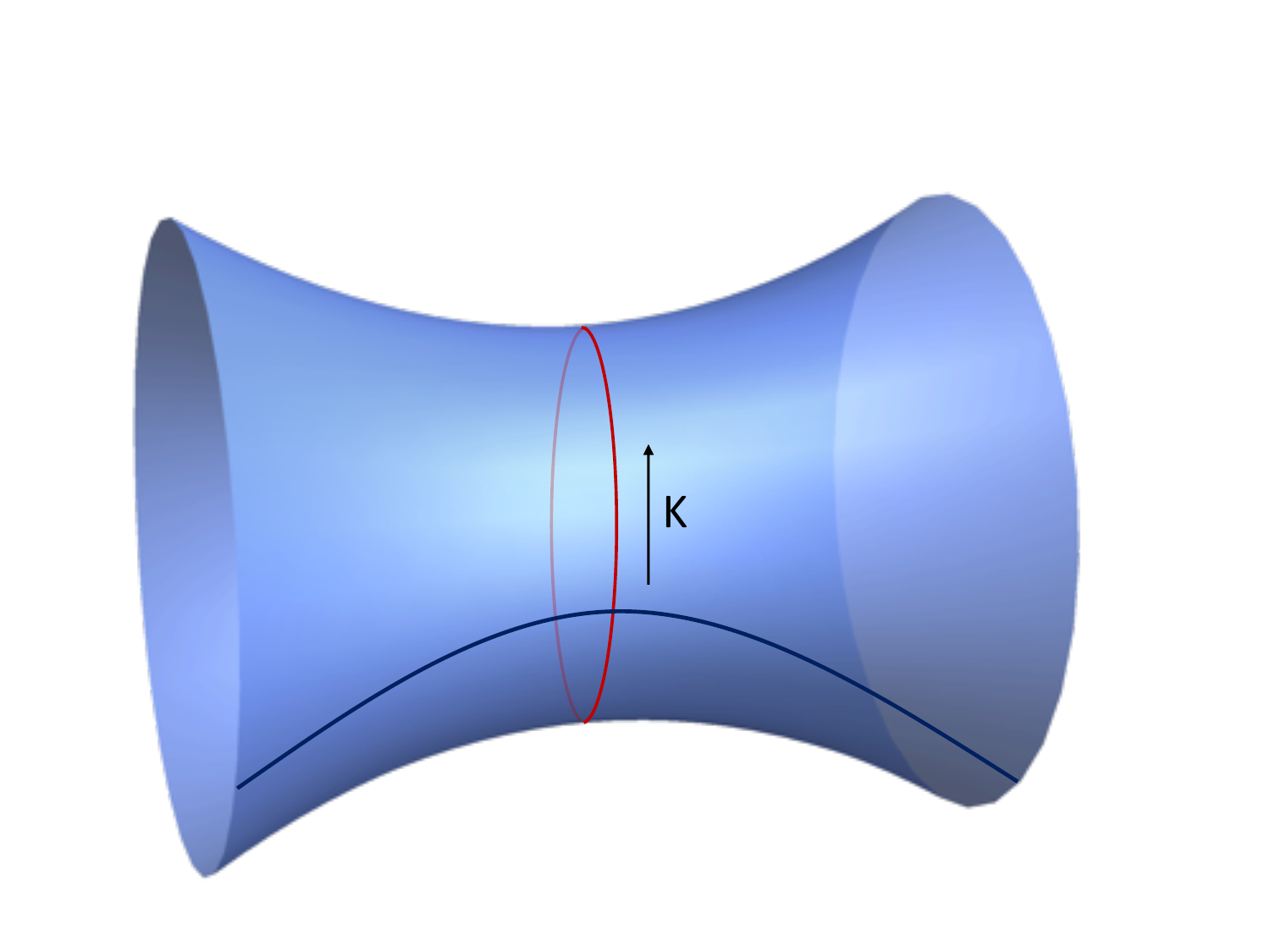}

\caption{Illustration of concepts of an Euclidean field theory on a two dimensional manifold describing an equilibrium state. The red line and the black line stand for time axis and space axis, respectively. The vector $K$ is a time-like Killing vector pointing along time direction, which origins from the static nature of equilibrium states. The time axis is further compactified to a thermal loop  (red line) so as to describe thermal physics, which in turn, yields two scalars. The first one is the local temperature $T(x)$, defined from length of thermal loop ($\beta(x)$), i.e., $T(x)\equiv 1/\beta(x)$.  The second one is the local chemical potential $\mu(x)$, defined from the Wilson loop of electromagnetic gauge fields $\oint A_0 dx^0_E$, i.e., $\mu(x)\equiv -\oint A_0dx^0_E/\beta(x)$.    \label{fig:local_temperature}}

\end{figure}
\section{Overview and summary of results}

Although transport is a non-equilibrium phenomenon, it is surprisingly simple that certain topological responses can be characterized from pure equilibrium aspects. Gaps between equilibrium and non-equilibrium quantities can be bridged by the Laughlin's argument~\cite{Laughlin:1981} as well as the Streda formula \cite{Streda_1982,prange1990springer, stone1992world}. 
To be more specific, Laughlin's argument tells us the quantum Hall response can be understood as the adiabatic response of a gapped ground state (equilibrium state property at zero temperature). Upon inserting a magnetic flux in a cylinder geometry, the adiabatic charge pumping process requires the anomalous edges to absorb the charge. This absorbing process is described by the anomaly inflow~\cite{callan1985npb}. For the other aspect, the Streda formula relates the Hall conductance of transport currents to thermodynamic property of magnetization in equilibrium. These two aspects are well-established for electric transports,  but the validity of the anomaly inflow aspect for thermal transports is still under debates. Problems are twofolds: On the one hand, previously proposed Laughlin's argument~\citep{nakai2016njp,nakai2017prb} for the thermal Hall effect requires the existence of bulk thermal Hall currents so as to absorb the edge quantum anomaly.  On the other hands, for gapped systems, there are hardly any bulk excitations, so as an entropy current the nonzero bulk thermal Hall currents are questionable.

Motivated by these aspects, we study the Euclidean field theory, which describes the bulk of a quantum Hall system at equilibrium. Especially, to address the anomaly matching problem we need to couple the system with background gravitational fields while maintain the sytem at equilibrium. Involving gravitational fields, equilibrium conditions are more subtle~\cite{luttinger1964pr, Cooper1997}: equilibrium is reached only when mechanical forces are balanced by statistical forces which stem from inhomogeneous distributions of charge or energy.  Interestingly, it turns out that these equilibrium conditions can be geometrically visualized in Fig.~\ref{fig:local_temperature}:  (i)~There exists a time-like Killing vector $K$ due to the static nature of equilibriums states. (ii)~The time axis along $K$ direction is compactified to a thermal loop (red line) so as to capture thermal physics. (iii)~Local temperature turns out to be the inverse of thermal loop length, and local chemical potential is the Wilson loop of electromagnetic fields along $K$ direction. For equilibrium states satisfying these equilibrium conditions, the generic formalism describing the physics at long length scale is the hydrostatics, or equivalently an Euclidean field theory equipped with a time-like Killing vector \cite{banerjee2012jhep, jensen2012prl}, where thermodynamic properties of the system are captured by a hydrostatic action from derivative expansions. Built on this setup, we will derive a hydrostatic action to linear power of derivatives [see Eq.~\eqref{eq:effective_action_magnetization}]. This action not only reproduces known results for electric transports, but more importantly, also clarifies the anomaly-inflow aspect of thermal transports. Namely, due to the previously overlooked back reaction of gravitational fields on temperature, our hydrostatic effective action can be recast as a topological $\theta$ term, so there is no bulk thermal Hall currents, and thus the Laughlin argument is invalid for the thermal Hall effect.


The paper is organized as follows:  In Sec.~\ref{subsec:einstein_relation}, we derive equilibrium conditions for the external field arising from the balancing between statistical forces and mechanical forces. We also obtain the conserved charge currents and energy currents from charge $U(1)$ and temporal translaton symmetry. 
In Sec.~\ref{subsec:energy_magnetization_Kubo}, we derive the hydrostatic effective action for magnetization as well as energy magnetization, whose relation to the microscopic linear response theory is studied in details in Sec.~\ref{sec: linear_response}. In Sec.~\ref{subsec:topo_action}, we show that our hydrostatic effective action for the thermal Hall effect can be recast as a topological $\theta$ term and thus there is no anomaly inflow as well as bulk thermal Hall currents at low temperature. Finally, in Sec.~\ref{subsec:topo_action}, we show that the thermal Hall conductance is quantized by global anomalies.

\section{Generalized Einstein's relation and conserved currents \label{subsec:einstein_relation}}
\Reply{As outlined in the previous section, we will focus on systems under time-independent external fields varying slowly in space, which reach their equilibrium 
when statistical forces from inhomogeneity of thermodynamic variables are balanced by mechanical forces. The equilibrium conditions for this balancing will be derived in this part from hydrostatics,}
where the static nature can be rephrased as the existence of translation symmetry along time-like Killing vector $K$. \Reply{For concreteness, let us consider a charge conserved thermal partition function ($Z\left[A_{\mu},\ e_{\mu}^{*a}\right]$) with Killing vector $K$ on a Euclidean spacetime manifold under slowly varying external vielbein $e_{\mu}^a$ and electromagnetic fields $A_{\mu}$.}
In this case,
the corresponding effective action $S_{\text{eff}}\left[A_{\mu},\ e_{\mu}^{*a}\right]=-i\ln Z$ can be organized in terms of derivative expansion. 
\Reply{In doing so we can first write down} all possible scalars invariant under symmetries of the system, which can be constructed from external fields and Killing vector. For clarity, we can denote these scalars as $s^{(n)}_i$ with the subscript $i$ labeling its power of derivative and the superscript $(n)$ for different scalars. 
From these scalars, the effective action can be organized as follows~\cite{jensen2012prl}, 
\begin{equation}
 S_{\text{eff}}= \int \sqrt{|\det g|} \{P[s_0]+\sum_{n}\sum_{i=1}\alpha_{i}^{(n)}[s_0] s_i^{(n)}\}\label{eq:effective_action_power_counting},
\end{equation}
where we have used the symbol $P$ for the zero-order term known as the internal pressure. \Reply{As we shall show later, local temperature and chemical potential manifest themselves as zeroth order scalars due to the temporal translational symmetry, whose equilibrium values are determined by the balancing between statistical forces and mechanical forces, and this yields equilibrium conditions. Combined with the $U(1)$ symmetry and the temporal translational symmetry, these equilibrium conditions enable us to define conserved charge currents as well as conserved energy currents.}


Let us now \Reply{construct the zeroth order scalars and relate them to local temperature and local chemical potential.} We first explicitly write down the metric of the Euclidean spacetime manifold
\begin{equation}
ds^{2}=g_{\mu\nu}dx^{\mu}\otimes dx^{\nu}=\left(e_{\mu}^{*a}dx_{E}^{\mu}\right)^{2},\label{eq:metric}
\end{equation}
where the subscript ``$E$'' denotes Euclidean spacetime and $e_{\mu}^{*a}$ is
the vielbein field. \XQ{Greek letters} $\mu,\ \nu$ and Latin \XQ{letters}
$a,\ b$ stand for Einstein indices and Lorentz indices, i.e., $\mu=0,\ 1,\ \dots$
and $a=0,\ 1,\ \dots$ We use $i,\ j$ and $I,\ J$ for spatial indices
of $\mu$ and $a$, respectively. If we recast $e_{\mu}^{*0}dx^{\mu}$ as $\left(1+\text{\ensuremath{\phi_{g}}}\right)dx^{0}+A_{gi}dx^{i}$,
then $\phi_{g}$ is Luttinger's fictitious gravitational field \citep{luttinger1964pr}
and $A_{gi}$ can be regarded as the gravitomagnetic fields. The imaginary
time axis is compactified to a circle so as to describe thermal effects \ZM{known as the thermal loop.}
In equilibrium, partition function should be time independent, so
we have a time-like Killing vector $K=\partial_{E0}$ (see Fig.~\ref{fig:local_temperature}) and its normalized
counterpart is 
\begin{equation}
u=\frac{1}{\sqrt{K^{2}}}\partial_{E0},
\label{eq:u-def}
\end{equation}
where $K^{2}\equiv K^{\mu}K^{\nu}g_{\mu\nu}=\left(1+\phi_{g}\right)^{2}$ \ZM{and $u$ thus points along the tangential direction of the thermal loop.}
For later convenience, we shall align $e_{\mu}^{*0}$ with $u_{\mu}$
hereafter, i.e., $e_{\mu}^{*0}dx^{\mu}=u_{\mu}dx^{\mu}$, which implies
that $e_{\mu}^{*I}u^{\mu}=0$ \XQ{and $e_{0}^{*0}=\sqrt{K^2}$}. 

Then, due to compactification of temporal
axis, in the presence of background vielbein $e_{\mu}^{*0}dx^{\mu}$
and $U\left(1\right)$ gauge fields $A_{\mu}$, we can define two
scalars \citep{banerjee2012jhep, jensen2012prl}, length of thermal loop, i.e., $\beta\left(x\right)$,
and Wilson loop along time direction, \Reply{which yield the local temperature as well as the chemical potential.}  The local temperature $T\left(x\right)$
is defined as the inverse of $\beta\left(x\right)$, i.e., 
\begin{equation}
T\left(x\right)\equiv\frac{1}{\int_{0}^{\frac{1}{T_{0}}}dx_{E}^{0}\sqrt{K^{2}}}\Reply{=\frac{T_0}{\sqrt{K^2}}=\frac{T_0}{e^{*0}_0}},\label{eq:temperature_local}
\end{equation}
where $\sqrt{K^{2}}$ is the induced metric of the thermal loop,  $x_{E}^{0}\in\left[0,\ 1/T_0\right]$ is the parametrization of thermal loops \footnote{Generally speaking, we shall parametrize them by $\sigma\in\left[0,\ 1/\tilde{T}_{0}\right]$, \ZM{where $\tilde{T}_0$ equals to $T_0$ when $\sigma=x_E^0$.}
The Killing vector $K$ becomes $\frac{d}{d\sigma}$, so length of
thermal loops becomes $\int_{0}^{1/\tilde{T}_{0}}d\sigma\sqrt{\frac{dx^{\mu}}{d\sigma}g_{\mu\nu}\frac{dx^{\nu}}{d\sigma}}$.
However, for simplicity, we have assumed that $x_{E}^{0}=\sigma$.
Because of this choice, we have $\mathcal{L}_{K}g_{\mu\nu}=\partial_{E0}g_{\mu\nu}=0$.
That is, $g_{\mu\nu}$ must be time-independent, so is $e_{\mu}^{*a}$.}, 
\Reply{and $T(x)$ satisfies the Tolman-Ehrenfest relation \citep{rovelli2011cgr} ($T\left(x\right)\sqrt{K^{2}}=\text{constant}$).}

The local chemical potential is defined as the temporal Wilson loop divided
by $\beta\left(x\right)$, i.e., 
\begin{equation}
\mu\left(x\right)\equiv-T\left(x\right)\int_{0}^{\frac{1}{T_{0}}}A_{0}dx_{E}^{0}\Reply{=-\frac{A_0}{\sqrt{K^2}}=-\frac{A_0}{e^{*0}_{0}}},\label{eq:chemical_potential_local}
\end{equation}
where $\int_{0}^{\beta_{0}}A_{0}dx_{E}^{0}$ is temporal Wilson loop \Reply{and the second equality is from the transverse gauge \citep{jensen2014jhep,banerjee2012jhep}: $\partial_{E0}A_0=0$.}
One can also define spin chemical potential as the Wilson loop
for spin connection ${\omega_{ab}}_{\mu}$ , i.e., $\frac{K^{\mu}\omega_{ab\mu}}{\sqrt{K^{2}}}$, but
as we shall show later, spin chemical potential should be set to zero
if we want to have a conserved energy current. 

\XQ{Eq.~(\ref{eq:temperature_local}) and Eq.~(\ref{eq:chemical_potential_local})
relate $T\left(x\right)$ and $\mu\left(x\right)$ to the gravitational
potential and electric potential}, respectively, so
in equilibrium, currents arising from inhomogenous particle (energy)
distribution are compensated by those from external electric fields
(torsional electric fields), \Reply{which are encoded in the time-independent
conditions, i.e., $0=\mathcal{L}_{K}g_{\mu\nu}=\mathcal{L}_{K}e_{\mu}^{*a}=\mathcal{L}_{K}A$
and they yield the generalized Einstein relations (for details, please
refer to Appendix \ref{sec:Einstein_relations}), }
\begin{equation}
T\nabla_{\nu}\frac{\mu}{T}-u^{\mu}F_{\mu\nu}=0,\label{eq:Einstein_1}
\end{equation}
and 
\begin{equation}
\frac{1}{T}\nabla_{\mu}T-{T^{a}}_{\sigma\mu}u_{a}u^{\sigma}=0,\label{eq:Einstein_2}
\end{equation}
where $F_{\mu\nu}\equiv\partial_{\mu}A_{\nu}-\partial_{\nu}A_{\mu}$
is the electromagnetic field strength tensor, $\frac{1}{2}{T^{a}}_{\sigma\mu}dx^{\sigma}\wedge dx^{\mu}\equiv de^{*a}+{\omega^{a}}_{b}\wedge e^{*b}$
is the torsion tensor and the spin chemical potential is set to
zero. These generalized Einstein's equations are valid even when we relax
the transverse gauge condition.

After obtaining these generalized Einstein relations, we turn to define
conserved charge current $\mathcal{J}^{\mu}$ as well as conserved
energy current $\mathcal{J}_{E}^{\mu}$, which \Reply{are from
$U(1)$ symmetry and the temporal translational symmetry, respectively.} From $U\left(1\right)$ symmetry, 
\begin{equation}
\mathcal{J}^{\mu}=\sqrt{\left|\det g\right|}j^{\mu},\ \partial_{\mu}\mathcal{J}^{\mu}=0
\end{equation}
where $j^{\mu}{\,\equiv\,}-\frac{1}{ \sqrt{\left|\det g\right|}}\frac{\delta S}{\delta A_{\mu}}$ satisfies $\frac{1}{\sqrt{\left|g\right|}}\partial_{\mu}\left(\sqrt{\left|\det g\right|}j^{\mu}\right)=0$. 
From temporal translational invariance
induced by $K$ (see Appendix \ref{subsec:Conserved_energy_currents_TT}
for details), one can define the energy currents as 
\begin{equation}
\mathcal{J}_{E}^{\mu}=\sqrt{\left|\det g\right|}\left[K^{a}\tau_{a}^{\mu}+\left(A_{\nu}K^{\nu}\right)j^{\mu}\right], \label{eq:energy_current}
\end{equation}
where $\tau^{\mu}_a\equiv -\frac{1}{\sqrt{|\det g|}}\frac{\delta S}{\delta e^{*a}_{\mu}}$ is energy-momentum tensor. This current is conserved if $K^{\mu}\omega_{ab\mu}$ vanishes \footnote{If spin chemical potential is not zero, then this current satisfies
\begin{eqnarray*}
 &  & \frac{1}{\sqrt{\left|\det g\right|}}\partial_{\mu}\sqrt{\left|\det g\right|}\left[\left(i_{K}e^{*b}\right)\tau_{b}^{\mu}+\left(i_{K}A\right)j^{\mu}-\left(i_{K}\omega\right)_{ab}S^{ab\mu}\right]\\
 & = & -\left(i_{K}{\omega^{c}}_{d}\right)e_{\mu}^{*d}\tau_{c}^{\mu}.
\end{eqnarray*}\protect\trick.
}, i.e., $\partial_{\mu}\mathcal{J}_{E}^{\mu}=0$. 
 For later
convenience, we shall set spin connections to zero hereafter so as
to have a conserved energy current and this is the teleparallel gravity
\citep{aldrovandi2013springer}.  In the absence of external electromagnetic fields and vielbeins, we have  $\mathcal{J}_E^{\mu}=\tau^{\mu}_a K^a -\mu j^\mu $. \Reply{Notice that conserved energy current here in the Euclidean theory is essentially the thermal current in the equilibrium states. This is special to the Euclidean theory at equilibrium and does \emph{not} imply thermal current to be conserved in real time evolution. However, in our paper, for consistency, we will keep the Euclidean theory terminology. Hence our later result of energy current and energy magnetization will correspond to thermal current and heat magnetization in literatures of the real time formalism.} 

\section{Hydrostatic effective action \label{subsec:energy_magnetization_Kubo}}
For the current responses, we can look for the derivative expansion of the first order and write the \XQ{general covariant form} of the action containing one derivative as:
\begin{equation}
S_{\text{eff}}^{(1)}=-\int m_{g,\ 0}\epsilon^{\mu\nu\rho}e_{\mu}^{*a}\partial_{\nu}e_{\rho }^{* b}\eta_{ab}-\int m_{N,\ 0}\epsilon^{\mu\nu\rho}u_{\mu}\partial_{\nu}A_{\rho},\label{eq:effective_action_first}
\end{equation}
where $m_{g,0}$ and $m_{N,0}$ are functions of zeroth order scalars, i.e. $m_{g,0}=m_{g,0}(\mu,T)$ and $m_{N,0}(\mu,T)$. \XQ{For simplicity, we assume the system to have emergent Lorentz symmetry such as in a Chern insulator, while it is straightforward to generalized to non-relativistic electrons. For the non-relativistic case, we need to treat space indices and time index differently but the main discussion of charge response and thermal response remains valid and only requires charge $U(1)$ and temporal translation symmetry. }\ZM{It is also worth pointing out that the celebrated Chern-Simons term $\frac{\nu_{H}}{4\pi}\int \epsilon^{\mu\nu\rho}A_\mu \partial_{\nu} A_\rho$ ($\nu_H \in \mathbb{Z}$) is contained in the second term of the action above, i.e.,   $-\int m_{N, 0}\epsilon^{\mu\nu\rho}u_\mu \partial_{\nu}A_\rho$ 
\footnote{We decompose gauge fields as $A=(i_{u}A) u +A_{\text{trans}}$, where $i_{u}$ is for the interior product with vector $u$. This implies that $i_{u}A_{\text{trans}}=i_{K}A_{\text{trans}}=0$.  Then, we have 
\begin{eqnarray*}
&&\int A\wedge dA\\
&=&\int \left[\left(i_{u}A\right)u+A_{\text{trans}}\right]\wedge dA\\
&=&\int \left[\left(i_{u}A\right)u\wedge dA+A\wedge dA_{\text{trans}}\right],
 \end{eqnarray*}
 where in the last line, we have used integral by parts. Because $i_{K}A_{\text{trans}}=0$ and $\mathcal{L}_{K}A_{\text{trans}}=0$, we have $i_{u}(dA_{\text{trans}})=0$, which means that $dA_{\text{trans}}$ is on the plane normal to $u$ and thus $A \wedge dA_{\text{trans}}=(i_{u}A) u\wedge  dA$. This shows that the Chern-Simons term can be recast as $\int m_{N, 0}u\wedge dA$ with $m_{N, 0}=-\frac{ \nu_{H} (i_u A)}{2\pi}=\frac{\nu_{H} \mu}{2\pi}$.}}. 
 
In order to bring more physical insights, we will justify the underlying physics of magnetization and energy magnetization for these coefficients in Eq.~\eqref{eq:effective_action_first} \Reply{by deriving this effective action from first principles and making connection with results in the Cooper-Halperin-Ruzin transport theory \citep{Cooper1997}}. 
 \Reply{The effective action is derived by coupling $\mathcal{J}^\mu$ and $\mathcal{J}^\mu_E$ to their probe fields, $A_\mu$ and $e^{*0}_i/e^{*0}_0$, i.e., 
 \begin{equation}
S_{\text{eff}}^{(1)}=-\int\left(\mathcal{J}_{E}^{i}-A_{0}\mathcal{J}^{i}\right)\left(\frac{1}{e_{0}^{*0}}e_{i}^{*0}\right)-\int\mathcal{J}^{i}A_{i},\label{eq:effective_action-1}
\end{equation}
where $\mathcal{J}^i_E-A_0\mathcal{J}^i\equiv \sqrt{|\det g|}K^a \tau^{\mu}_a$ couples to $e^{*0}_i/e^{*0}_0$ and zero components of currents are not written down due to the time independent condition. This time independent condition implies that  $\partial_{\mu}\mathcal{J}_{\left(E\right)}^{\mu}=\partial_{i}\mathcal{J}_{\left(E\right)}^{i}=0$, so these conservation laws are solved by $\mathcal{J}^{i}=\partial_{j}m_{N}^{ij}$ and $\mathcal{J}_{E}^{i}=\partial_{j}m_{g}^{ij}$,
with skew symmetric $m_{N}^{ij}$ and $m_{g}^{ij}$} known as magnetization and energy magnetization \cite{Cooper1997, qin2011prl,zhang2020prb}, respectively. As we can see from our hydrostatic theory, (energy) magnetization currents are equilibrium currents in the presence of inhomogeneous background fields that do not participate in transport~\cite{Cooper1997,Smrcka:1977}. Especially, the energy magnetization current is important to substract to give the correct thermal Hall response theory~\cite{qin2011prl}. 
\ZM{In $(2+1)$-dimensions, magnetization and energy magnetization can be further recast as $m_{N}^{ij}=\epsilon^{ij0}m_N$ and $m_{g}^{ij}=\epsilon^{ij0}m_g$. These solutions of currents $\mathcal{J}^{i}$ and $\mathcal{J}_{E}^{i}$
are of first-order dependence in the derivative expansion Eq.~(\ref{eq:effective_action_power_counting}), so they should be encoded in the action in Eq.~\eqref{eq:effective_action_first}. This can be straightforwardly appreciated by recasting the action in Eq.~\eqref{eq:effective_action-1}  in terms of magentization and energy magnetization.}

\begin{equation}
S_{\text{eff}}^{(1)}=-\int m_{g,\ 0}\epsilon^{\mu\nu\rho}e_{\mu}^{*0}\partial_{\nu}e_{\rho}^{*0}-\int m_{N,\ 0}\epsilon^{\mu\nu\rho}e_{\mu}^{*0}\partial_{\nu}A_{\rho},\label{eq:effective_action_magnetization}
\end{equation}
where $m_{N,\ 0}$ and
$m_{g,\ 0}$ are defined as 

\begin{subequations}
\label{eq:energy_number_magnetization}
\begin{eqnarray}
 m_{N}&\equiv&\sqrt{K^{2}}m_{N,\ 0}, \\
 m_{g} & \equiv & K^{2}m_{g,\ 0}+\sqrt{K^{2}}\left(i_{K}A\right)m_{N,\ 0}\nonumber \\
 & = & K^{2}\left(m_{g,\ 0}-\mu m_{N,\ 0}\right),
\end{eqnarray}
\end{subequations}
and this is one of our main results. Notice that $K^{2}\equiv K^{\mu}K^{\nu}g_{\mu\nu}=1+\phi_{g}$, so Eq.~\eqref{eq:energy_number_magnetization} reproduces
scaling relations suggested in Ref.~\citep{Cooper1997}. 
\Reply{It is worth pointing out that the action in Eq.~\eqref{eq:effective_action_magnetization} does match the one in Eq.~\eqref{eq:effective_action_first},} because in our metric, $e_{0}^{*I}=0,\ e_{\mu}^{*I}u^\mu=0$ and thus $\int\epsilon^{\mu\nu\rho}e_{\mu}^{*I}\partial_{\nu}e_{\rho}^{*I}=0$.
In general choice of coordinates,
for zero spin connections, the effective action Eq.~\eqref{eq:effective_action_magnetization} is covariantly generalized to Eq.~\eqref{eq:effective_action_first}. 

\Reply{Finally, let us highlight two comments about our effective action:} First, the coefficient
$m_{g,\ 0}$ in Ref.~\citep{hughes2011prl} depends on ultra-violet
cut-off, 
which in our formalism has a physical meaning of zero-temperature energy magnetization and enables us to better understand these UV divergences. 
\XQ{Second, our results of magnetization and energy magnetization reproduce previous study in Ref.~\citep{Cooper1997}.} Namely, in terms of $\phi_{g}$, we can determine the functional form of the magnetization and energy magnetization to be $m_{N}^{ij}=\epsilon^{0ij}\left(1+\phi_{g}\right)m_{N,0}\left(\mu,\ T\right)$
and $m_{g}^{ij}=\epsilon^{0ij}\left(1+\phi_{g}\right)\left[\left(1+\phi_{g}\right)m_{g,\ 0}\left(\mu,\ T\right)+A_{0}m_{N,\ 0}\right]$,
which are the scaling relations suggested in Ref.~\citep{Cooper1997}. 


\section{Effective action and linear response theory}\label{sec: linear_response}
Our effective action describes the macroscopic property of a system at a (in)homogeneous equilibrium state. Now in this section, we obtain the equations for the (energy) magnetization by connecting our effective action to microscopics, and highlight general constraints of (energy) magnetization for gapped systems. Starting with our effective action, these equations do \emph{not} depend on microscopic details other than the symmetry of the system. 
\Reply{More specifically, the (energy) magnetization currents must match in calculations by (i) variating our hydrostatic effective action and (ii) linear response theory from a microscopic theory. 
For the former, our hydrostatic action yields }
\begin{subequations}
\label{eq:variation_action}
\begin{eqnarray}
\mathcal{J}^{i}&=&\epsilon^{ij0}\left(-\frac{\partial m_{N,\ 0}}{\partial\mu}\right)\partial_{j}A_{0}\nonumber \\
&&+\epsilon^{ij0}\left(m_{N,\ 0}-\mu\frac{\partial m_{N,\ 0}}{\partial\mu}-T\frac{\partial m_{N,\ 0}}{\partial T}\right)\partial_{j}e^{*0}_{0},\nonumber\\
&&\\
\mathcal{J}_{E}^{i}&=&\epsilon^{ij0}\left(-\frac{\partial m_{Q,\ 0}}{\partial\mu}\right)\partial_{j}A_{0}\nonumber \\
&&+\epsilon^{ij0}\left(2m_{Q,\ 0}-\mu\frac{\partial m_{Q,\ 0}}{\partial\mu}-T\frac{\partial m_{Q,\ 0}}{\partial T}\right)\partial_{j}e^{*0}_{0},\nonumber\\
&&
\end{eqnarray}
\end{subequations}
where as we analyzed above, $m_{N,0}$ and $m_{g,0}$ are functions of local chemical potential $\mu$ and temperature $T$. $m_{Q,\ 0}\equiv m_{g,\ 0}-\mu m_{N,\ 0}$ \ZM{ and we have used Eqs. (\ref{eq:temperature_local},\ \ref{eq:chemical_potential_local}) to rewrite gradient of local temperature and chemical potential gradient in terms of gradient of $e^{*0}_0$ and $A_0$. Eq. (\ref{eq:variation_action}) can also be derived from the definition of magnetization currents, i.e. $\mathcal{J}^i=\epsilon^{ij0}\partial_j m_N$ and $\mathcal{J}_E^i=\epsilon^{ij0}\partial_j m_g$.}

For the latter, \ZM{perturbative expansions, or equivalently Feynman diagrams, lead to
\begin{subequations}
\begin{eqnarray}
\mathcal{J}^{i}\left(q\right)&=&-\langle \mathcal{J}^{i}\left(q\right)\mathcal{J}^{0}\left(-q\right)\rangle \delta A_{0}\left(q\right)\nonumber \\
&&-\langle \mathcal{J}^{i}\left(q\right)\sqrt{|\det{g}|}\tau_{0}^{0}\left(-q\right)\rangle \delta e^{*0}_{0}(q),\\
\mathcal{J}_{E}^{i}&=&-\langle\mathcal{J}_{E}^{i}\left(q\right)\mathcal{J}^{0}\left(-q\right)\rangle \delta A_{0}(q)\nonumber\\
&&-\langle\mathcal{J}_{E}^{i}\left(q\right)\sqrt{|\det{g}|}\tau_{0}^{0}\left(-q\right)\rangle \delta e^{*0}_{0}(q),
\end{eqnarray}
\end{subequations}
where we have only kept terms from linear perturbations.}  By comparing
results from these two approaches, we can obtain a set of equations for $m_{N,0}$ and $m_{Q,0}$
\begin{subequations}
\label{kubo_formula}
\begin{eqnarray}
\frac{\partial m_{N,\ 0}}{\partial\mu}&=&\frac{i}{2}\epsilon_{ki0}\partial_{q_{k}}\langle \mathcal{J}^{i}\left(q\right)\mathcal{J}^{0}\left(-q\right)\rangle,\nonumber\\
&&\label{eq:Kubo_1}\\
\left(m_{N,\ 0}-T\frac{\partial m_{N,\ 0}}{\partial T}\right)&=&-\frac{i}{2}\epsilon_{ki0}\partial_{q_{k}}\langle \mathcal{J}^{i}\left(q\right)\mathcal{J}_{E}^{0}\left(-q\right)\rangle,\nonumber \\
&&\label{eq:Kubo_2}\\
\frac{\partial m_{Q,\ 0}}{\partial\mu}&=&\frac{i}{2}\epsilon_{ki0}\partial_{q_{k}}\langle\mathcal{J}_{E}^{i}\left(q\right)\mathcal{J}^{0}\left(-q\right)\rangle,\nonumber \\
&&\label{eq:Kubo_3}\\
\left(2m_{Q,\ 0}-T\frac{\partial m_{Q,\ 0}}{\partial T}\right)&=&-\frac{i}{2}\epsilon_{ki0}\partial_{q_{k}}\langle\mathcal{J}_{E}^{i}\left(q\right)\mathcal{J}_{E}^{0}\left(-q\right)\rangle,\nonumber\\
&&\label{eq:Kubo_4}
\end{eqnarray}
\end{subequations}
which reproduce
results in Ref.~\citep{qin2011prl}.  These are first-order differential equations, so we can obtain
$m_{g,\ 0}$ and $m_{N,\ 0}$ unambiguously only when references states
are provided. \Reply{Still, these differential equation provide valuable insights on constraints for magnetization in a gapped system.} \Reply{Most importantly, in a system with gap $\Delta$, $\mathcal{J}_E$ is expected to be exponentially suppressed, i.e., $e^{-\beta \Delta}$,  because there are hardly any excitations in the bulk and thus entropy are exponentially suppressed. }
\Reply{When combining this exponential suppression with Eqs.~(\ref{eq:Kubo_1}, \ref{eq:Kubo_2}), we have $m_{N,\ 0}=\frac{\nu_H \mu}{2\pi}+c_2 T$, where $c_2$ is a constant and $\nu_H \in \mathbb{Z}$. The $\frac{\nu_H \mu}{2\pi}$ term is a well-known result from the integer quantum Hall effect and the $c_2 T$ term is from Eq.~(\ref{eq:Kubo_2}) by setting terms on right-handed side to zero.} As for Eqs.~(\ref{eq:Kubo_3}, \ref{eq:Kubo_4})
 with terms of the right-handed equal zero,
their solution is $m_{Q,\ 0}=c_{1}T^{2}$ and $c_{1}$ is a constant. \ZM{By putting these results together, we have $m_{g, 0}=m_{Q, 0}+\mu m_{N, 0}=c_1 T^2 +c_2 \mu T +\frac{\nu_{H}\mu^2}{2\pi}$.} \Reply{Two comments are in order:} First, $c_1$ and $c_2$ can not be determined perturbatively from Eqs.~\eqref{kubo_formula} given above, which as we shall shown in the next section, is because that they are rooted in boundary modes. Second, the $\mu^2$ term in $m_{g,\ 0}$ can give rise to another torsional Chern-Simons term, i.e., $S= -\int \frac{\mu^2 \nu_{H}}{2\pi} \epsilon^{\mu \nu \rho}e^{*a}_{\mu}\partial_{\nu}e^{*b}_{\rho}\eta_{ab}$. It is interesting to notice that cut-offs in the Hughes-Leigh-Fradkin
parity-odd action \citep{hughes2011prl} are replaced by chemical
potential and its quantization inherits from the integer quantum Hall
effect. \Reply{Finally, it is worth pointing out that in experimental systems, there can exist gapless phonons which yield finite contributions to $\mathcal{J}_E$  \cite{vinkler-aviv2018prx,ye2018prl}.}

\section{$m_{Q,\ 0}=c_{1}T^{2}$, bulk-edge correspondence and its topological
meaning \label{subsec:topo_action}}

As we have discussed above, $m_{Q,\ 0}$ is expected to be $c_{1}T^{2}$ for an insulator \ZM{and thus $m_{g, 0}=c_1 T^2 +c_2 \mu T+\frac{\nu_{H} \mu^2}{2\pi}$, where both $c_1$ and $c_2$ can not be fixed from bulk perturbative calculations.}
In this part, we shall turn to the torsional Chern-Simons term with
$m_{g,\ 0}=c_{1}T^{2}+c_2\mu T$ and explore its topological meanings as well
as bulk-edge correspondence. \ZM{The $c_2$ term can be recast as a boundary term \footnote{The corresponding derivation is given as follow
\begin{eqnarray*}
&&\int\left(c_{2}\mu T\right)\epsilon^{\mu\nu\rho}e_{\mu}^{*0}\partial_{\nu}e_{\rho}^{*0}+\int\left(c_{2}T\right)\epsilon^{\mu\nu\rho}e_{\mu}^{*0}\partial_{\nu}A_{\rho}\\
&=&-c_{2}\int\frac{A_{0}T_{0}}{e_{0}^{*0}}\epsilon^{0ij}\partial_{i}e_{j}^{*0}-c_{2}\int\frac{A_{0}T_{0}}{\left(e_{0}^{*0}\right)^{2}}\epsilon^{ij0}e_{i}^{*0}\partial_{j}e_{0}^{*0}\\
&&+c_{2}\int T_{0}\epsilon^{0ij}\partial_{i}A_{j}+c_{2}\int T_{0}\epsilon^{ij0}\frac{e_{i}^{*0}}{e_{0}^{*0}}\partial_{j}A_{0}\\
&=&+c_{2}\int T_{0}\epsilon^{0ij}\partial_{i}\left(A_{j}-\frac{e_{j}^{*0}}{e_{0}^{*0}}A_{0}\right),
\end{eqnarray*}
which turns out to be a topological theta term as well.}, but by direct calculations of edge chiral fermions, one can find that $c_2=0$, so we shall focus on the $c_1$ term hereafter~\footnote{This can be appreciated as follow. Consider edge chiral fermions with dispersion $\mathcal{E}=\mathcal{E}(p)$ and velocity $v = \frac{\partial \mathcal{E}}{\partial p}$, where $\mathcal{E}$ takes values from $-\infty$ to $+\infty$ due to its chiral nature. The corresponding current can be calculated as follow
  \begin{eqnarray*}
j^{1}&=&\int_{-\infty}^{+\infty} \frac{d p}{2 \pi}\frac{\partial{\mathcal{E}}}{\partial p} [\frac{1}{e^{\beta_0 (\mathcal{E}-\mu)}+1}-\theta(-\mathcal{E})] \\
&=&s \int_{-\infty}^{+\infty}  \frac{d\mathcal{E}}{2 \pi} [\frac{1}{e^{\beta_0 (\mathcal{E}-\mu)}+1}-\theta(-\mathcal{E})] \\
&=&s \frac{\mu}{2\pi},
\end{eqnarray*} 
where $s=\pm 1$ is from chiralities and $\theta(-\mathcal{E})$ is a regulator used to subtract contributions from Dirac sea. This shows that edge currents are independent of temperature, so we have $c_2=0$.}}. \Reply{As we will see, this term can be recast as a topological $\theta$ term and thus endow the thermal Hall conductance with topological meaning.} 

\Reply{We shall first reveal the topological nature of this $c_1 T^2$ and then show how to extract its physical information.} To this end, we first study its robustness under small perturbations: under variations of $e_{\mu}^{*0}$, the $c_1 T^2$ term is invariant up to a boundary term, i.e., $-\int\epsilon^{\mu\nu\rho}\partial_{\nu}\left(c_{1}T^{2}e_{\mu}^{*0}\delta e_{\rho}^{*0}\right)$, 
so this term is robust against bulk perturbations and
it can not be obtained from bulk perturbative calculations. This is because this $c_1 T^2$ term is secretly a topological $\theta$ term, and we can rewrite it as
\footnote{The corresponding derivation is given as follow 
\begin{eqnarray*}
&&\int T^{2}\epsilon^{\mu\nu\rho}e_{\mu}^{*0}\partial_{\nu}e_{\rho}^{*0}\\&=&\int T_{0}^{2}\epsilon^{0ij}\left(\frac{1}{e_{0}^{*0}}\right)^{2}\left(e_{0}^{*0}\partial_{i}e_{j}^{*0}+e_{i}^{*0}\partial_{j}e_{0}^{*0}\right)\\&=&\int T_{0}^{2}\epsilon^{0ij}\left[\frac{1}{e_{0}^{*0}}\partial_{i}e_{j}^{*0}-\frac{1}{\left(e_{0}^{*0}\right)^{2}}e_{j}^{*0}\partial_{i}e_{0}^{*0}\right]\\&=&\int T_{0}^{2}\epsilon^{0ij}\partial_{i}\left(e_{j}^{*0}/e_{0}^{*0}\right).
\end{eqnarray*}\protect\trick.}, 
\begin{equation}
-\int c_{1}T_{0}^{2}\epsilon^{0ij}\partial_{i}\left(e_{j}^{*0}/e_{0}^{*0}\right),\label{action_theta}
\end{equation}
\Reply{where $e^{*0}_j/e^{*0}_0$ is the emergent Kaluza-Klein gauge fields associated with temporal translation as we can see:} (i)~Under local spatial translation,  $\left(e_{j}^{*0}/e_{0}^{*0}\right)$
transforms like a conventional spatial one-form. (ii)~Under local temporal
translation, i.e., $\delta x^{\mu}=\xi^{0}\left(\boldsymbol{x}\right)\delta_{0}^{\mu}$,
we have $\delta\left(e_{i}^{*0}/e_{0}^{*0}\right)=\partial_{i}\xi^{0}$, \Reply{which is an effective $U(1)$ gauge transformation  due to the identification $\xi^{0}\simeq\xi^{0}+\frac{1}{T_0}$.}
Despite robustness of this topological theta term against bulk variations, we can still extract its physical information by considering two \XQ{adjacent} materials with different value of $c_{1}$. 
\XQ{Around the boundary, the system is inhomogeneous and $c_1$ can develop dependence on the coordinate across the boundary. To be more concrete, we assume the two materials locate at $y>0$ and $y<0$ with a smooth boundary, and we model $c_1$ as a function of $y$ interpolating between these two materials. The corresponding effective edge theory is thus
given as
\begin{equation}
-\int d^2x dy c_{1}(y)T_{0}^{2}\epsilon^{0ij}\partial_{i}\left(e_{j}^{*0}/e_{0}^{*0}\right) =\int\tilde{c}_{1}T^{2}e_{0}^{*0}e_{1}^{*0}d^2x,  
\label{eq:interpolation}
\end{equation}
where $\tilde{c}_{1}\equiv -[c_{1}(+\infty)-c_{1}(-\infty)]$.} The lowest-order approximation of this edge action reproduces results in Refs.~\citep{nakai2016njp,nakai2017prb}.
However, the corresponding physical meanings are different in a significant way. Compared to results in Refs.~\citep{nakai2016njp,nakai2017prb},
our bulk effective action is invariant under temporal coordinate transformations,
so its effective edge action can not be obtained from anomaly matching
and there are no bulk energy currents \XQ{derived from} our bulk action. 

\Reply{One can further reads off boundary energy-momentum tensor from our effective edge theory and then study the edge conservation laws. Namely,  $\tau_{0}^{1}\PRLequal -\tilde{c}_{1}T^{2}\frac{1}{\sqrt{\left|\det g\right|}}e_{0}^{*0}$
and $\tau_{0}^{0}\PRLequal \tilde{c}_{1}T^{2}\frac{1}{\sqrt{\left|\det g\right|}}e_{1}^{*0}$}, and it is rather interesting to find that it satisfies the
corresponding Noether theorem \XQ{in the presence of torsion (see Appendix \ref{sec:Noether_current})}, i.e., $\frac{1}{\sqrt{\left|\det g\right|}}\partial_{\nu}\left(\sqrt{\left|\det g\right|}\tau_{a}^{\nu}\right)-e_{a}^{\mu}\left(\tau_{b}^{\nu}{T^{b}}_{\mu\nu}\right)=0$.
Hence, this term is not from perturbative diffeomorphism anomaly at
edges, which is significantly different from its zero-temperature
counterparts~\citep{hughes2013prd,onkar2014prd}. \Reply{As for the edge energy current}, by definition,
it is $\mathcal{J}_{E,\ \text{boundary}}^{i}=\sqrt{\left|\det g\right|}K^{a}\tau_{a}^{\mu}=-\tilde{c}_{1}T_{0}^{2}$, so
it is clearly conserved. \XQ{Compared to the Chern-Simons action for integer quantum quantum Hall effect,
this $c_{1}T^{2}$ torsional Chern-Simons term fails to cause anomaly currents flowing from
bulks to edges.} It was claimed in Ref.~\citep{nakai2016njp} that a bulk thermal Hall current
is needed to compensate for this boundary anomalies. We would like to
stress that energy currents $\mathcal{J}_{E}^{\mu}$ are different
from $\tau_{0}^{\mu}$. Hence, one can not interpret $\frac{1}{\sqrt{\left|\det g\right|}}\partial_{\nu}\left(\sqrt{\left|\det g\right|}\tau_{a}^{\nu}\right)=e_{a}^{\mu}\left(\tau_{b}^{\nu}{T^{b}}_{\mu\nu}\right)$
as energy-current non-conservation, \Reply{and the Laughlin's argument is not applicable for the thermal Hall effect in gapped systems. Still, energy pumping between boundaries through the bulk is possible if there exists gapless modes, e.g. phonons  \cite{vinkler-aviv2018prx, ye2018prl}.} 

Due to the robustness of the \XQ{temperature-squared} term in our effective action, a natural question is
how we can calculate this term from microscopic model and fix the
coefficient $c_{1}$ \XQ{in the bulk of a homogeneous material.} 
\Reply{The answer is $m_{g,\ 0}$ can be uniquely fixed only when boundary conditions, or reference states are given, which is because Eqs.~\eqref{kubo_formula} for $m_{g,\ 0}$ are first-order differential equations.}
For example, we have condition:  $m_{g,\ 0}|_{\mu\rightarrow-\infty}=0$ by setting energy magnetization in vacuum to zero. \Reply{Also,
we can impose condition $\frac{\partial m_{g,\ 0}}{\partial T}|_{T\rightarrow\infty}=0$ because all quasiparticles are excited in the $T\rightarrow\infty$ limit.} For a demonstration of ths approach on $\left(2+1\right)$-dimensional massive Dirac fermions, interested readers are referred to Appendix~\ref{subsec:THE_2+1} for details. 

 In reality, we can implement these boundary conditions or reference states by putting two different materials with different $\mu$ or $T$ \XQ{adjacent}
to each other, for example, $\mu=0$ and $\mu=-\infty$, respectively. Assuming $\mu$   smoothly
interpolates between these two materials, \XQ{in the same spirit with Eq.~\eqref{eq:interpolation}}, our effective action $\int c_{1}T^{2}\epsilon^{\mu\nu\rho}e_{\mu}^{*0}\partial_{\nu}e_{\rho}^{*0}$
manifest itself as energy currents flowing in the interface determined by the difference of $c_1$ in the bulk of two materials. Since the variation of the action in the bulk is zero, only the edge physics determined by
the difference of $c_{1}$ is observable, so the torsional effective
action $\int c_{1} T^{2}\epsilon^{\mu\nu\rho}e_{\mu}^{*0}\partial_{\nu}e_{\rho}^{*0}$
is topological in a relative sense~\citep{kapustin2020prb}.

\begin{figure}
\includegraphics[scale=0.25]{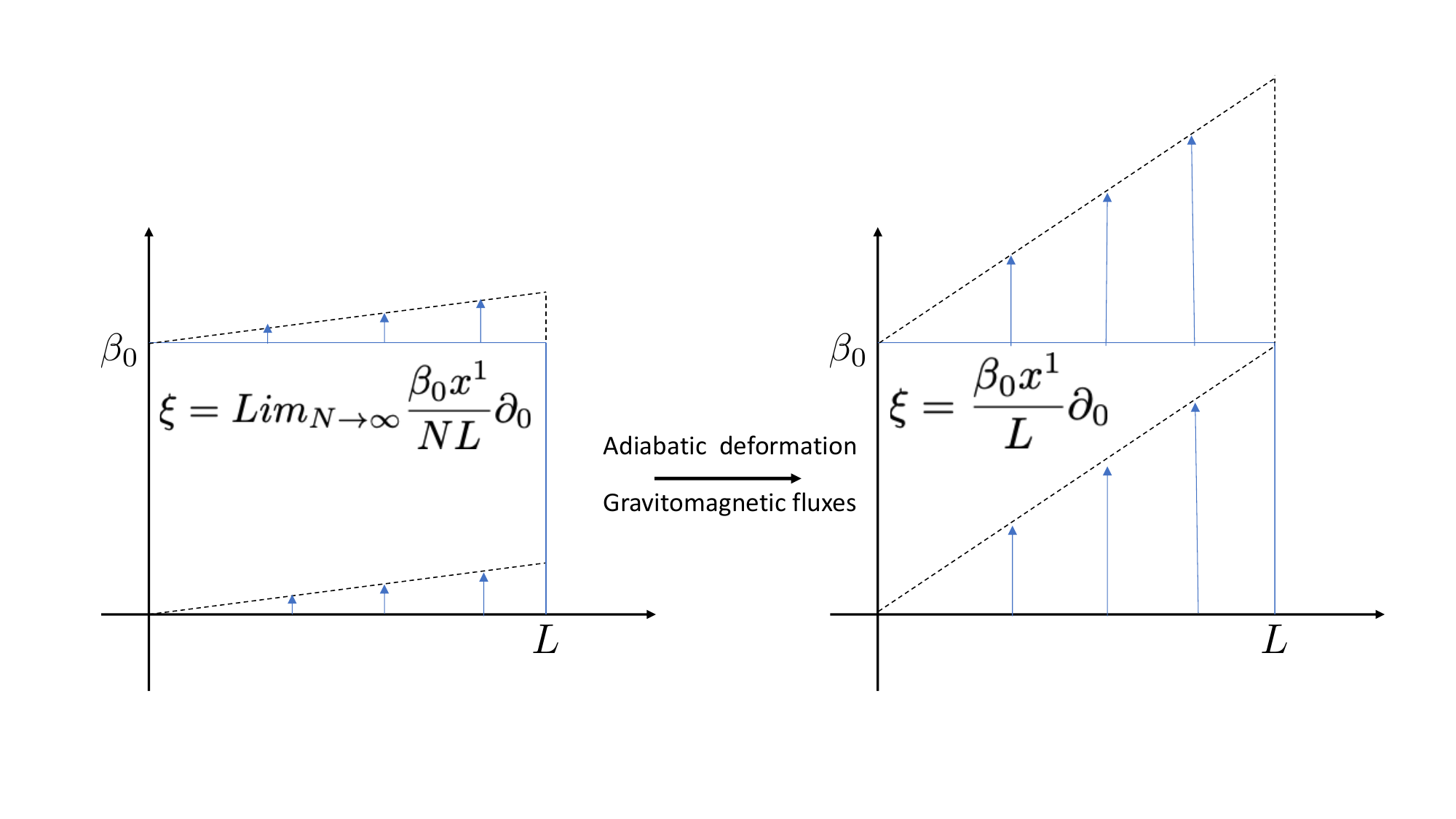}

\caption{Gravitomagnetic fluxes and modular transformation. Left panel: an infinitesimal deformation of torus along temporal direction. Right panel: a modular transformation that maps a torus onto itself.   \label{fig:gravitomagnetic_flux_modular}}

\end{figure}

\section{$m_{Q,\ 0}=c_{1}T^{2}$ and global anomalies }\label{sec:global_anomaly}

After revealing the topological meaning of torsional Chern-Simons
terms, a natural question is whether $c_{1}$ is quantized after considering the scale invariance of the edge theory similarly as in Ref.~\citep{nakai2017prb} and fixing the vacuum energy magnetization to be 0. We shall explore
this by studying boundary energy-momentum tensors \Reply{from the point view of perturbative calculations and non-perturbative global anomalies (modular transformation)}. 

To this end, consider the boundary between a topologically non-trivial material at $y>0$ and a vacuum at $y<0$, which traps right-handed chiral fermions at $y=0$ if the Chern number equals one for the topological material. Now we demonstrate that we can fix the $c_1$ for the topological material by studying edge chiral fermions.
For example, we can directly compare the boundary action Eq.~\eqref{eq:interpolation} with microscopic calculation of the energy momentum tensor of right-handed chiral fermions (see Appendix \ref{sec:energy_momentum_tensor} for details): 
\begin{equation}
\langle\tau_{0}^{1}\rangle=\left(\tilde{\Lambda}^{2}+\frac{\pi}{12}T_{0}^{2}\right)+\mathcal{O}\left[\left(\delta e_{\mu}^{*a}\right)^{2}\right].\label{edge_em_tensor}
\end{equation}
Here, $\tilde{\Lambda}^{2}$ is the ultra-violet cut-off and its specific value
depends on regularization schemes. 
\Reply{For example, by using the dimensional regularization so as to preserve the scale
invariance in edge theories, $\tilde{\Lambda}^2=0$, while $\tilde{\Lambda}^2\neq0$ by using the hard-off regularization.} \Reply{From Eq.~\eqref{edge_em_tensor}, and compare with our edge action Eq.~\eqref{eq:interpolation}, we can further conclude that $c_1=\frac{\pi}{12}$ for this topological material. }

Alternatively, we can fix the value of $c_1$ non-perturbatively by compacifying the spatial dimension and considering the global anomaly of the edge theory on a torus. One reason of doing so is to compare with Ref.~\citep{nakai2017prb}. In addition, this approach will not refer to microscopic details of the edge theory and therefore is a stronger argument. Now the idea is to connect global anomaly of the partition function under the modular transformation of the torus to the field theory response to an inserted gravitomagnetic flux as can be described in our boundary action. 
The compactification to torus is done by 
identifying spacetime coordinates in the
following way: $\left(x_{E}^{0},\ x^{1}\right)\sim\left(x_{E}^{0}+\beta_{0},\ x^{1}\right)\sim\left(x_{E}^{0},\ x^{1}+L\right)$, where $\beta_0 \equiv \frac{1}{T_0}$. Boundary conditions for fermion are (anti-) periodic
along (temporal) spatial direction.

We then insert a gravitomagnetic flux quanta to deform this spacetime torus and mimick the modular transformation as shown in Fig.~\ref{fig:gravitomagnetic_flux_modular}. This flux quanta insertion is implemented as the sum of a seris of infintesimal gravitomagnetic fluxes (e.g $\oint \delta e^{*0}_i dx^i =\frac{\beta_0}{N}, \ N\to\infty$), where the latter can be geometrically represented as an infinitesimal deformation generated by vector fields $\xi^{\mu}\partial_{\mu}=\frac{\beta_{0}}{N L}x^{1}\partial_{E0}$ (see Fig.~\ref{fig:gravitomagnetic_flux_modular}).  This deformation changes our boundary action by  $\delta S=\int \tau^{0}_1 \delta e^{^*0}_1$ with $\delta e^{*0}_1=\partial_{x^1}\xi^0=\frac{\beta_0}{NL}$. After this process, the torus is mapped to itself (see Fig.~\ref{fig:gravitomagnetic_flux_modular}), but with coordinate basis changed, which is known as the modular transformation \cite{francesco1997springer}. The ensuing action transformation is $\delta S =\int \tau^{1}_0 \delta e^{*0}_{1}$ with $\delta e^{*0}_1=\sum_N\partial_{x^1}\xi^0=\frac{\beta_0}{L}$ for a gravitomagnetic flux quanta.  

\Reply{Now we can match results from global anomalies and responses to gravitomagnetic flux quanta.} On the one hand, the modular transformation changes the partition functions for chiral fermions by a phase factor \citep{hori2003ams,francesco1997springer,nakai2017prb},
i.e., $Z\rightarrow e^{i\frac{\pi}{12}}Z$. On the other hand, responses to a gravitomagnetic flux quanta is $\delta S=\int \tau^1_0 \frac{\beta_0}{L}$, where $\tau^1_0 =c_1 T^2\frac{1}{\sqrt{|\det g|}} e^{*0}_0$ is from our boundary action and $\frac{\beta_0}{L}$ is from gravitomagnetic flux quanta.  Combining these, we find $e^{i\frac{\pi}{12}}= e^{i\int c_{1}T_{0}^{2}\frac{\beta_0}{L}d^{2}x}$,
and thus again we conclude $c_{1}=\frac{\pi}{12}$, which shows that $c_{1}$ is quantized by global anomalies. 
Since the $c_{1}$ is from the topological state of Chern number $+1$, we can conclude that for the topological state of Chern number $+1$, the energy magnetization is $\frac{\pi}{12}T^2$, and from the thermal generalization of the Streda formula \citep{nomura2012prl}, the thermal Hall conductivity is $\kappa_H=-\frac{\pi}{6}T$.

\section{Conclusions}

In summary, \XQ{we have derived the general effective hydrostatic action for gapped quantum matter coupling to telleparallel gravity. The action up to linear order of the derivative expansion recovers the charge and energy currents response and covariantly generalizes the thermoelectric transport theory. The linear order in derivative terms include a torsional Chern-Simons term, with its physical meaning as the energy magnetization. For a gapped system, there can exist a temperature-squared torsional Chern-Simons term in our hydrostatic effective action, which is topological and its quantization inherits from boundary global
gravitational anomalies. In contrast to previous literatures discussing the torional Chern-Simons term, in our theory there is no boundary diffeomorphism
anomalies and bulk inflow currents. In addition, we have derived various
properties for the magnetization as well as the energy magnetization from our effective action.} 


\section{Acknowledgement}

The authors wish to thank Barry Bradlyn, Jing-Yuan Chen, Haoyu Guo, Kristan Jensen, Biao Lian,  Laimei Nie, Atsuo Shitade and Mike Stone for useful
discussions, especially Mike for invaluable feedback. Z.-M. H was not
directly supported by any funding agency, but this work would not
be possible without resources provided by the Department of Physics
at the University of Illinois at Urbana-Champaign. B. H. was supported
by ERC Starting Grant No. 678795 TopInSy. X.-Q. S acknowledges support
from the Gordon and Betty Moore Foundations EPiQS Initiative through
Grant GBMF8691.

\appendix

\section{Derivation of Noether's current from diffeomorphism \label{sec:Noether_current}}

In this part, we shall derive the Noether current arising from general
coordinate invariance.

For a given action $S$, we define charge currents as 
\begin{equation}
j^{\mu}\equiv-\frac{1}{\sqrt{\left|\det g\right|}}\frac{\delta S}{\delta A_{\mu}},
\end{equation}
energy-momentum tensors as 
\begin{equation}
\tau_{a}^{\mu}\equiv-\frac{1}{\sqrt{\left|\det g\right|}}\frac{\delta S}{\delta e_{\mu}^{*a}},
\end{equation}
and spin currents as
\begin{equation}
S^{ab\mu}\equiv\frac{1}{\sqrt{\left|\det g\right|}}\frac{\delta S}{\delta\omega_{ab\mu}},
\end{equation}
where $A_{\mu}$, $e_{\mu}^{*a}$ and $\omega_{ab\mu}$ are $U\left(1\right)$
gauge fields, vielbeins and spin connections, respectively. In addition,
we define $\nabla_{\mu}$ as the total covariant derivative acting
on both Einstein indices $\mu$ and Lorentz indices $a$, which contains
both spin connections $\omega_{ab\mu}$ and affine connections ${\Gamma^{\mu}}_{\nu\rho}$.
$D_{\mu}$ is used for covariant derivative and it only contains spin connections.

We consider a coordinate transformation generated by vector $\xi^{\mu}$.
The variation of fields $e_{\mu}^{*a},\ \omega_{ab\mu}$ and $A_{\mu}$
are 
\begin{equation}
\delta e^{*a}=\mathcal{L}_{\xi}e^{*a}=i_{\xi}T^{a}+D\xi^{a}-\left(i_{\xi}{\omega^{a}}_{b}\right)e^{*b},
\end{equation}
\begin{equation}
\delta{\omega^{a}}_{b}=\mathcal{L}_{\xi}{\omega^{a}}_{b}=i_{\xi}{\Omega^{a}}_{b}+D\left(i_{\xi}{\omega^{a}}_{b}\right),
\end{equation}
and 
\begin{equation}
\delta A=\mathcal{L}_{\xi}A=i_{\xi}F+di_{\xi}A,
\end{equation}
where $i_{\xi}$ denotes interior products and ${\Omega^{a}}_{b} \equiv d{\omega^{a}}_b+{(\omega\wedge \omega)^a}_b$ is curvature.
Variations of action are
\begin{widetext}
\begin{eqnarray}
\delta S & = & \int\sqrt{\left|\det g\right|}[-\left(-\frac{1}{\sqrt{\left|\det g\right|}}\frac{\delta S}{\delta e_{\nu}^{*a}}\right)\delta e_{\nu}^{*a}\nonumber \\
 &  & +\frac{1}{\sqrt{\left|\det g\right|}}\frac{\delta S}{\delta\omega_{ab\nu}}\delta\omega_{ab\nu}-\left(-\frac{1}{\sqrt{\left|\det g\right|}}\frac{\delta S}{\delta A_{\nu}}\right)\delta A_{\nu}]\nonumber \\
 & = & \int\sqrt{\left|\det g\right|}[\xi^{\mu}\left(-\tau_{a}^{\nu}{T^{a}}_{\mu\nu}+S^{ab\nu}\Omega_{ab\mu\nu}-j^{\nu}F_{\mu\nu}\right)+\left(\nabla_{\nu}+{T^{\rho}}_{\nu\rho}\right)\tau_{a}^{\nu}\xi^{a}]\nonumber \\
 &  & +\int\sqrt{\left|\det g\right|}\left(i_{\xi}{\omega^{a}}_{b}\right)\left[e_{\nu}^{*b}\tau_{a}^{\nu}-\frac{1}{\sqrt{\left|\det g\right|}}D_{\mu}\left(\sqrt{\left|\det g\right|}{S_{a}}^{b\mu}\right)\right]\nonumber \\
 &  & +\int\sqrt{\left|\det g\right|}\left(i_{\xi}A\right)\left(\frac{1}{\sqrt{\left|\det g\right|}}\partial_{\mu}\sqrt{\left|\det g\right|}j^{\mu}\right).
\end{eqnarray}
Because of $U\left(1\right)$ symmetry, we have $\frac{1}{\sqrt{\left|\det g\right|}}\partial_{\mu}\sqrt{\left|\det g\right|}j^{\mu}=0,$
and the Noether currents from general coordinate invariance is 
\begin{equation}
\left(\nabla_{\nu}+{T^{\rho}}_{\nu\rho}\right)\tau_{a}^{\nu}-e_{a}^{\mu}\left(\tau_{b}^{\nu}{T^{b}}_{\mu\nu}-S^{cd\nu}\Omega_{cd\mu\nu}+j^{\nu}F_{\mu\nu}\right)=-{\omega^{c}}_{da}\left[e_{\nu}^{*d}\tau_{c}^{\nu}-\frac{1}{\sqrt{\left|\det g\right|}}D_{\mu}\left(\sqrt{\left|\det g\right|}{S_{c}}^{d\mu}\right)\right].\label{eq:general_coordiante_invariance}
\end{equation}
If there exists internal rotational symmetry among indices $a$, then
one can prove that 
\begin{equation}
{\omega^a}_{bc} [e_{\nu}^{*b}\tau_{a}^{\nu}-\frac{1}{\sqrt{\left|\det g\right|}}D_{\mu}\left(\sqrt{\left|\det g\right|}{S_{a}}^{b\mu}\right)]=0,
 \end{equation}
and the Noether current in Eq.~(\ref{eq:general_coordiante_invariance})
becomes 
\begin{equation}
\left(\nabla_{\nu}+{T^{\rho}}_{\nu\rho}\right)\tau_{a}^{\nu}-e_{a}^{\mu}\left(\tau_{b}^{\nu}{T^{b}}_{\mu\nu}-S^{cd\nu}\Omega_{cd\mu\nu}+j^{\nu}F_{\mu\nu}\right)=0,
\end{equation}
which matches results in Ref.~\citep{bradlyn2015prb} and Ref.~\citep{onkar2014prd}.
\end{widetext}

\section{Derivation of Eq.~(\ref{eq:Einstein_1}), Eq.~(\ref{eq:Einstein_2})
and conserved energy current \label{sec:Einstein_relations}}

In this part, we shall give a detailed derivation of the generalized
Einstein relation from equilibrium conditions, i.e., $\mathcal{L}_{K}\left(\dots\right)=0$,
where $\left(\dots\right)$ stands for external fields, including
$A_{\mu},\ g_{\mu\nu}$ and so on. In addition, the derivations of
conserved energy currents are also presented in details.

\subsection{Derivation of Eq.~(\ref{eq:Einstein_1})}

We impose following equilibrium condition 
\begin{equation}
\mathcal{L}_{K}\left(A+d\theta\right)=0,\label{eq:equlibrium}
\end{equation}
where $d\theta$ is a gauge transformation and this says that $A$
satisfies $\mathcal{L}_{K}A=0$ up to a gauge transformation. Eq.
(\ref{eq:equlibrium}) can be recast as

\begin{equation}
0=\mathcal{L}_{K}\left(A+d\theta\right)=i_{K}dA+d\left(i_{K}A+i_{K}d\theta\right).
\end{equation}
Following conventions in Ref.~\citep{jensen2014jhep}, we define $\Lambda_{K} \equiv i_{K}d\theta$
and chemical potential $-T_{0}\frac{\mu}{T\left(x\right)}=i_{K}A+\Lambda_{K}$.
Correspondingly, we have found, 
\begin{equation}
0=K^{\mu}F_{\mu\nu}-T_{0}\partial_{\nu}\frac{\mu}{T}=u^{\mu}F_{\mu\nu}-T\partial_{\nu}\frac{\mu}{T}.
\end{equation}
 For simplicity, one usually use the transverse gauge condition, i.e.,
$i_{K}d\theta=0$. This says that a gauge-fixing condition is imposed
to get rid of time-dependence in gauge transformation parameter $\theta$.

\subsection{Derivation of Eq.~(\ref{eq:Einstein_2})}

Similar to Eq.~(\ref{eq:Einstein_1}), we can derive Eq.~(\ref{eq:Einstein_2})
by imposing the following condition 
\begin{equation}
\mathcal{L}_{K}u_{\mu}=0.
\end{equation}
To be more specific, $\mathcal{L}_{K}u_{\mu}$ can be calculated as
follow

\begin{eqnarray}
0 & = & \frac{1}{\sqrt{K^{2}}}\mathcal{L}_{K}u_{\mu}\nonumber \\
 & = & u^{\nu}\nabla_{\nu}u_{\mu}+\frac{1}{\sqrt{K^{2}}}\nabla_{\mu}K^{\nu}u_{\nu}\nonumber \\
 &  & -\frac{1}{\sqrt{-K^{2}}}{T^{\sigma}}_{\mu\rho}u_{\sigma}K^{\rho}\nonumber \\
 & = & u^{\nu}\nabla_{\nu}u_{\mu}-\frac{1}{T}\nabla_{\mu}T-{T^{\sigma}}_{\mu\rho}u_{\sigma}u^{\rho}.
\end{eqnarray}
where we have used

\begin{eqnarray}
\mathcal{L}_{K}u_{\mu} & = & K^{\nu}\left(\partial_{\nu}u_{\mu}-{\Gamma^{\alpha}}_{\mu\nu}u_{\alpha}\right)\nonumber \\
 &  & +\left(\partial_{\mu}K^{\nu}+{\Gamma^{\nu}}_{\alpha\mu}K^{\alpha}\right)u_{\nu}-{T^{\nu}}_{\mu\alpha}K^{\alpha}u_{\nu}\nonumber \\
 & = & K^{\nu}\nabla_{\nu}u_{\mu}+\nabla_{\mu}K^{\nu}u_{\nu}-{T^{\nu}}_{\mu\alpha}u_{\nu}K^{\alpha}.
\end{eqnarray}
Noticed that $u_{\mu}dx^{\mu}=e_{\mu}^{*0}dx^{\mu}$, so we have
\begin{equation}
\nabla_{\nu}u_{\mu}=\partial_{\nu}e_{\mu}^{*0}-{\Gamma^{\alpha}}_{\mu\nu}e_{\alpha}^{*0}=-{\omega^{0}}_{b\nu}e_{\mu}^{*b},
\end{equation}
and 
\begin{equation}
K^{\nu}\nabla_{\nu}u_{\mu}=-{\omega^{0}}_{b\nu}e_{\mu}^{*b}K^{\nu},
\end{equation}
where we have used the following identity 
\begin{equation}
\partial_{\mu}e_{\nu}^{*a}-{\Gamma^{\alpha}}_{\nu\mu}e_{\alpha}^{*a}+{\omega^{a}}_{b\mu}e_{\nu}^{*b}=0.
\end{equation}
For metric 
\begin{equation}
ds^{2}=\left(e_{\mu}^{*a}dx^{\mu}\right)\otimes\left(e_{\nu}^{*b}dx^{\nu}\right)\eta_{ab},
\end{equation}
 with $e_{\mu}^{*0}=u_{\mu}$, we have $e_{\mu}^{*I}u^{\mu}=0$. This
means that $K^{\nu}\nabla_{\nu}u_{\mu}=-{\left(i_{K}\omega\right)^{0}}_{\mu}$
and thus 
\begin{equation}
\frac{1}{T}\nabla_{\mu}T-{T^{a}}_{\sigma\mu}u_{a}u^{\sigma}+\left(i_{u}\omega\right)_{0\mu}=0,
\end{equation}
If the spin chemical potential is set to zero, the equation above becomes 
\begin{equation}
\frac{1}{T}\nabla_{\mu}T-{T^{a}}_{\sigma\mu}u_{a}u^{\sigma}=0.
\end{equation}

\subsection{Conserved energy current}

\subsubsection{Conserved energy currents from diffeomorphism and temporal translation
symmetry}

The Noether currents from diffeomorphism is (see Appendix \ref{sec:Noether_current}
for details)

\begin{eqnarray}
 &  & \frac{1}{\sqrt{\left|\det g\right|}}D_{\nu}\left(\sqrt{\left|\det g\right|}\tau_{a}^{\nu}\right)\nonumber \\
 &&-e_{a}^{\mu}\left(\tau_{b}^{\nu}{T^{b}}_{\mu\nu}-S^{cd\nu}\Omega_{cd\mu\nu}+j^{\nu}F_{\mu\nu}\right)\nonumber\\
 & = & -{\omega^{c}}_{da}\left[e_{\nu}^{*d}\tau_{c}^{\nu}-\frac{1}{\sqrt{\left|\det g\right|}}D_{\mu}\left(\sqrt{\left|\det g\right|}{S_{c}}^{d\mu}\right)\right].\label{eq:general_coordiante_invariance-1}\nonumber\\
\end{eqnarray}
where $D_{\mu}$ is the covariant derivative with only spin connections,
but not $\Gamma$. By using $0=\mathcal{L}_{K}e_{\mu}^{*a}=\mathcal{L}_{K}A_{\mu}=\mathcal{L}_{K}\omega_{ab\mu}$,
we have following identities

\begin{eqnarray}
 &  & \frac{1}{\sqrt{\left|\det g\right|}}\partial_{\mu}\left[\sqrt{\left|\det g\right|}\left(i_{K}e^{*b}\right)\tau_{a}^{\mu}\right]\nonumber \\
 & = & -\left(i_{K}T^{a}\right)_{\mu}\tau_{a}^{\mu}+\left(i_{K}e^{*a}\right)\frac{1}{\sqrt{\left|\det g\right|}}D_{\mu}\left(\sqrt{\left|\det g\right|}\tau_{a}^{\mu}\right),\nonumber\\
\end{eqnarray}
\begin{eqnarray}
 &  & \frac{1}{\sqrt{\left|\det g\right|}}\partial_{\mu}\left[\sqrt{\left|\det g\right|}\left(i_{K}A\right)j^{\mu}\right]\nonumber \\
 & = & -F_{\nu\mu}K^{\nu}j^{\mu},\nonumber\\
\end{eqnarray}
and 
\begin{eqnarray}
 &  & \frac{1}{\sqrt{\left|\det g\right|}}\partial_{\mu}\left[\sqrt{\left|\det g\right|}\left(i_{K}\omega\right)_{ab}S^{ab\mu}\right]\nonumber \\
 & = & -\left(i_{K}\Omega_{ab}\right)_{\mu}S^{ab\mu}+\left(i_{K}\omega\right)_{ab}\frac{1}{\sqrt{\left|\det g\right|}}D_{\mu}\left(\sqrt{\left|\det g\right|}S^{ab\mu}\right).\nonumber\\
\end{eqnarray}
They leads to 
\begin{widetext}
\begin{eqnarray}
 &  & \frac{1}{\sqrt{\left|\det g\right|}}\partial_{\mu}\sqrt{\left|\det g\right|}\left[\left(i_{K}e^{*a}\right)\tau_{a}^{\mu}+\sqrt{\left|\det g\right|}\left(i_{K}A\right)j^{\mu}\right]\nonumber \\
 & = & -S^{\mu ab}\left(i_{K}\Omega_{ab}\right)_{\mu}-{\omega^{c}}_{da}\left[e_{\mu}^{*d}\tau_{c}^{\mu}-\frac{1}{\sqrt{\left|\det g\right|}}D_{\mu}\left(\sqrt{\left|\det g\right|}{S_{c}}^{d\mu}\right)\right],
\end{eqnarray}
or 
\begin{eqnarray}
 &  & \frac{1}{\sqrt{\left|\det g\right|}}\partial_{\mu}\sqrt{\left|\det g\right|}\left[\left(i_{K}e^{*b}\right)\tau_{b}^{\mu}+\left(i_{K}A\right)j^{\mu}-\left(i_{K}\omega\right)_{ab}S^{ab\mu}\right]\nonumber \\
 & = & -\left(i_{K}{\omega^{c}}_{d}\right)e_{\mu}^{*d}\tau_{c}^{\mu},
\end{eqnarray}
\end{widetext}
which is conserved if we set the background spin connections, or curvature
to zero. We thus define the conserved energy current as 
\begin{eqnarray}
\mathcal{J}_{E}^{\mu} & = & \sqrt{\left|\det g\right|}\left[\left(i_{K}e^{*a}\right)\tau_{a}^{\mu}+\left(i_{K}A\right)j^{\mu}\right].
\end{eqnarray}

\subsubsection{Conserved energy currents from temporal translation symmetry \label{subsec:Conserved_energy_currents_TT}}

The energy current $\mathcal{J}_{E}^{\mu}$ defined above can be understood
from global translation symmetry directly. In equilibrium, under temporal
translations, we have 
\begin{equation}
0=\delta e_{\mu}^{*a}=\delta A_{\mu},
\end{equation}
and 
\begin{equation}
\delta\psi=K^{\mu}\partial_{\mu}\psi,
\end{equation}
where we have set $\omega_{ab\mu}=0$. Correspondingly, the Noether current associated with temporal translation is 
\begin{equation}
\mathcal{J}_{E}^{\mu}=-\left(\frac{\partial \sqrt{|\det g|}\mathcal{L}}{\partial\partial_{\mu}\psi}\delta\psi+\text{h.\ c. }-K^{\mu}\sqrt{|\det g|}\mathcal{L}\right),
\end{equation}
where $\mathcal{L}$ is Lagrangian density and $g$ is metric.
Assuming that action $S$ depends on vielbeins through $\det g$
and $D_a=e_{a}^{\mu}\left(\partial_{\mu}+A_{\mu}\right)$, we can recast
$\mathcal{J}_{E}^{\mu}$ as 
\begin{eqnarray}
\mathcal{J}_{E}^{\mu} & = & - \left[\frac{\partial \sqrt{|\det g|}\mathcal{L}}{\partial\partial_{\mu}\psi}K^{\alpha}\left(\partial_{\alpha}+iA_{\alpha}\right)\psi+\text{h.\ c. }\right]\nonumber \\
 &  & +K^{\mu}\sqrt{|\det g|}\mathcal{L} +\left(K^{\alpha}A_{\alpha}\right)\left(i\frac{\partial \sqrt{|\det g|}\mathcal{L}}{\partial \partial_{\mu}\psi}\psi+\text{h.\ c.\ }\right)\nonumber \\
 & = & -\left[\frac{\partial \sqrt{|\det g|}\mathcal{L}}{\partial D_{\alpha}\psi}\frac{\partial D_{\alpha}\psi}{\partial e_{\mu}^{*a}}K^{a}+\text{h.\ c. }\right] \nonumber \\
 &  & +K^{\mu}\sqrt{|\det g|}\mathcal{L}+\sqrt{\left|\det g\right|}\left(K^{\alpha}A_{\alpha}\right)j^{\mu}\nonumber \\
 & = & -\frac{\delta S}{\delta e_{\mu}^{*a}}K^{a}+\sqrt{\left|\det g\right|}\left(K^{\alpha}A_{\alpha}\right)j^{\mu}\nonumber \\
 & = & \sqrt{\left|\det g\right|}\left[\left(i_{K}e^{*a}\right)\tau_{a}^{\mu}+\left(i_{K}A\right)j^{\mu}\right],\nonumber\\
\end{eqnarray}
where the $U\left(1\right)$ Noether current is defined as $j^{\mu}\equiv\frac{1}{\sqrt{\left|\det g\right|}}\left(i\frac{\partial S}{\partial\partial_{\mu}\psi}\psi+\text{h.\ c}\right)$.
This matches with our results obtained before.

\section{Energy magnetizations for $\left(2+1\right)$-dimensional Dirac fermions \label{subsec:THE_2+1}}

In this part, we shall provide a detailed derivation of energy magnetizations
for $\left(2+1\right)$-dimensional massive Dirac fermions. 

\begin{figure}
\includegraphics[scale=0.5]{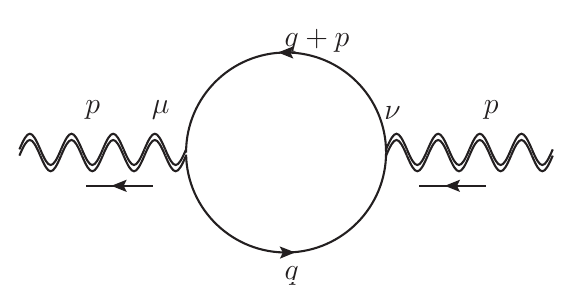}\caption{Feynman diagram for $\langle\tilde{\tau}_{a}^{\mu}\left(-p\right)\tilde{\tau}_{b}^{\nu}\left(p\right)\rangle$
. $\tilde{\tau}_{a}^{\mu}$ is defined as $\tilde{\tau}_{a}^{\mu}\equiv\tau_{a}^{\mu}+\delta_{a}^{\mu}\mathcal{L}$,
where the $-\delta_{a}^{\mu}\mathcal{L}$ term in $\tau_{a}^{\mu}$
is subtracted. Double wavy lines stands for external vielbeins. \label{fig:Feynman_diagram}}
\end{figure}

We consider the following action for $\left(2+1\right)$-dimensional
Dirac fermions in the Minkowski spacetime
\begin{equation}
S=\int d^{3}x\sqrt{\left|\det g\right|}\left[\frac{1}{2}\left(\bar{\psi}\gamma^{a}ie_{a}^{\mu}\partial_{\mu}\psi-\bar{\psi}i\overleftarrow{\partial}_{\mu}\gamma^{a}e_{a}^{\mu}\psi\right)-m\bar{\psi}\psi\right],
\end{equation}
where both chemical potentials and external electromagnetic fields
are set to zero. Energy-momentum tensors are

\begin{equation}
\tau_{b}^{\nu}=\frac{1}{2}\left(\bar{\psi}\gamma^{\nu}i\partial_{b}\psi+\text{h.c}\right)-\delta_{b}^{\nu}\mathcal{L}.,
\end{equation}
and we have set the spin connection to zero. In addition, we are most
interested in the (thermal) Hall effect, so the $\delta_{b}^{\nu}\mathcal{L}$
term is neglected hereafter and we define $\tilde{\tau}_{b}^{\nu}=\tau_{b}^{\nu}+\delta_{b}^{\nu}\mathcal{L}$.
Then, values of the Feynman diagram in Fig. \ref{fig:Feynman_diagram}
in equilibrium implies that the response energy current is 
\begin{equation}
\mathcal{J}_{E,\ \text{equ}}^{\mu}=-\frac{C_{\text{equ}}m}{4\pi}\epsilon^{\mu\alpha\nu}\partial_{\alpha}e_{\nu}^{*0},\label{eq:thermal_current}
\end{equation}
where 
\begin{equation}
C_{\text{equ}}=T_{0}\left[\frac{\left|m\right|}{T_{0}}\tanh\left(\frac{1}{2}\frac{\left|m\right|}{T_{0}}\right)-\frac{\Lambda}{T_{0}}\tanh\left(\frac{1}{2}\frac{\Lambda}{T_{0}}\right)\right].
\end{equation}
This means that the term on the right-handed side of Eq.~(\ref{eq:Kubo_4}) equals to
$-\frac{m}{4\pi}C_{\text{equ}}$. By solving Eq.~(\ref{eq:Kubo_4}),
one can obtain the energy magnetization, i.e., 
\begin{eqnarray}
m_{g,\ 0} & = & c_{1}T_{0}^{2}-\frac{mT_{0}}{8\pi}\{\left(\frac{\left|m\right|}{T_{0}}-\frac{\Lambda}{T_{0}}\right)+4[\frac{T_{0}}{\Lambda}\text{Li}_{2}\left(-e^{-\frac{\Lambda}{T}}\right)\nonumber \\
 &  & -\frac{T_{0}}{\left|m\right|}\text{Li}_{2}\left(-e^{-\frac{\left|m\right|}{T_{0}}}\right)]+4\ln\left(\frac{1+e^{-\left|m\right|/T_{0}}}{1+e^{-\Lambda/T_{0}}}\right)\},\nonumber\\
\end{eqnarray}
where $c_{1}$ can not be determined by solving Eq.~(\ref{eq:Kubo_4}).
In the low temperature limit, i.e., $T\rightarrow0$, we have $m_{g,\ 0}\simeq-\frac{1}{8\pi}\left(m\left|m\right|-m\Lambda\right)$, so
the effective action in Eq.~(\ref{eq:effective_action_magnetization})
becomes $-\frac{m\left(\left|m\right|-\Lambda\right)}{8\pi}\int\epsilon^{\mu\nu\rho}e_{\mu}^{*0}\partial_{\nu}e_{\rho}^{*0}$
, which matches the torsional Chern-Simons term obtained in Ref.~\citep{hughes2011prl,hughes2013prd}.
Similarly, at finite temperature, the $c_{1}T_{0}^{2}$ term suggests
that there exists a thermal torsional Chern-Simons term, i.e., $-c_{1}\int T_{0}^{2}\epsilon^{\mu\nu\rho}e_{\mu}^{*0}\partial_{\nu}e_{\rho}^{*0}$.

Now let us determine the value of $c_1$ by taking the high temperature limit as reference states. Consider an ultra-violet complete model at high temperature,
we expect all quasiparticles are excited, so $m_{g,\ 0}$ should be
temperature independent. In this limit, $m_{g,\ 0}=c_{1}T_{0}^{2}-\left[\frac{m\left(\left|m\right|-\Lambda\right)}{8\pi}+\frac{\pi\text{sign}\left(m\right)}{24}T_{0}^{2}\right]$,
where $m\ll\Lambda$, which combined with the temperature-independent condition, yields
\begin{equation}
c_{1}=\frac{\pi\text{sign}\left(m\right)}{24}T_{0}^{2}.
\end{equation}
This means that we have fixed $c_1$
by imposing physical conditions, even though it can not be determined from perturbative calculations of Feynman diagrams. From this point of view, $c_{1}T_{0}^{2}$
looks like counterterms arise from ultra-violet physics.

In summary,  $m_{g,\ 0}$ in the low-temperature limit is
\begin{equation}
m_{g,\ 0}=\text{sign}(m)\frac{\pi}{24}T_{0}^{2}-\frac{m\left(\left|m\right|-\Lambda\right)}{8\pi},\label{eq:energy_magnetization}
\end{equation}
and the ensuing effective action is
\begin{eqnarray}
S_{\text{eff}} & = & -\int\left[\frac{\pi\text{sign}\left(m\right)}{24}T_{0}^{2}+\frac{m\left(\Lambda-\left|m\right|\right)}{8\pi}\right]\epsilon^{\mu\nu\rho}e_{\mu}^{*0}\partial_{\nu}e_{\rho}^{*0}\nonumber \\
 &  & +\mathcal{O}\left[\left(\phi_{g}\right)^{2}\right].
\end{eqnarray}

\section{Calculations of energy-momentum tensor $\langle\tau_{0}^{1}\rangle$ in $(1+1)$-dimensional spacetime
\label{sec:energy_momentum_tensor}}

In this part, we shall present calculations of energy-momentum
tensor in flat spacetime. By definition, we have
\begin{eqnarray}
 &  & \langle\tau_{0}^{1}\rangle \nonumber \\
 & = & \langle\bar{\psi}\gamma^{1}p_{0}\left(\frac{1+s\gamma_{5}}{2}\right)\psi\rangle\nonumber \\
 & = & \frac{1}{2}\int\frac{dp_{1}}{2\pi}\left(\sum_{n}\frac{1}{\beta_{0}}\right)\frac{\left(p_{0}\right)^{2}}{p^{2}}\text{tr}\left(s\gamma^{0}\gamma^{1}\gamma_{5}\right)\nonumber \\
 & = & s \int\frac{dp_{1}}{2\pi}\left(\sum_{n}\frac{1}{\beta_{0}}\right)\frac{\left(p_{0}\right)^{2}}{p^{2}},
\end{eqnarray}
where $s=\pm1$ is for chiralities of Weyl fermions, gamma matrices are defined as  $\gamma^{0}=\sigma^1$, $\gamma^1=i\sigma^2$ and $\gamma^5=-\sigma^3$.

\subsubsection{Hard-cutoff regularizations}

Now we shall calculate the integral above by using hard-cut-off
regularizations, i.e., 

\begin{eqnarray}
 &  & \int\frac{dp_{1}}{2\pi}\left(\sum_{n}\frac{1}{\beta_{0}}\right)\frac{\left(i\omega_{n}\right)^{2}}{\left(i\omega_{n}\right)^{2}-p_{1}^{2}}\nonumber \\
 & = & \int\frac{dp_{1}}{2\pi}\left[-\frac{1}{2}\epsilon n_{F}\left(-\epsilon\right)+\frac{1}{2}\epsilon n_{F}\left(\epsilon\right)\right]\nonumber\\
 & = & 2\pi\int\frac{d\epsilon}{2\pi}\frac{\epsilon}{2\pi}n_{F}\left(\epsilon\right)\nonumber \\
 & = & 2\pi\left\{ \int_{-\infty}^{+\infty}\frac{d\epsilon}{2\pi}\frac{\epsilon}{2\pi}\left[n_{F}\left(\epsilon\right)-\theta\left(-\epsilon\right)\right]+\int_{-4\pi\tilde{\Lambda}}^{+\infty}\frac{d\epsilon}{2\pi}\frac{\epsilon}{2\pi}\theta\left(-\epsilon\right)\right\} \nonumber\\
 & = & \left(\frac{\pi}{12\beta_{0}^{2}}+\tilde{\Lambda}^{2}\right),
\end{eqnarray}
where $\epsilon\equiv\left|p_{1}\right|$ is the energy for Weyl fermions,
$n_{F}\equiv\frac{1}{e^{\beta_{0}\epsilon}+1}$ is the Fermi-Dirac
distribution function and $\tilde{\Lambda}$ is a cut-off. 

\subsubsection{Dimensional regularization }

If we use dimensional regularization instead, the $\tilde{\Lambda}^{2}$
term vanishes, i.e.,

\begin{eqnarray}
 &  & \int\frac{dp_{1}}{2\pi}\left(\sum_{n}\frac{1}{\beta_{0}}\right)\frac{\left(i\omega_{n}\right)^{2}}{\left(i\omega_{n}\right)^{2}-p_{1}^{2}}\nonumber \\
 & = & \frac{1}{2\beta_{0}}\sum_{n}\left|\omega_{n}\right|\nonumber \\
 & = & \pi\beta_{0}^{-2}\left(1+3+5+\dots\right)\nonumber \\
 & = & \frac{\pi}{12\beta_{0}^{2}},
\end{eqnarray}
where in the second line, we have integrated over $p_{1}$ by using
dimensional regularization. In the last line, we have used $\left(1+3+5\dots\right)=\frac{1}{12}$,
which is because $\sum_{n=1}^{+\infty}n=-\frac{1}{12}$ and $2\sum_{n=1}^{+\infty}n+\left(1+3+\dots\right)=-\frac{1}{12}$.
Note that $\sum_{n}\left|\omega_{n}\right|$ is the vacuum energy
of fermions.

\subsubsection{Results }
In summary, we have found 
\begin{equation}
\tau_{0}^{1}=s\left(\frac{\pi}{12\beta_{0}^{2}}+\tilde{\Lambda}^{2}\right).
\end{equation}

\end{document}